\documentclass[12pt]{article}
\usepackage{graphicx}

\newcommand\pubdate{\today}

\newcommand\pubnumber{CMS CR-2011/163}

\def\Title#1{\begin{center} {\Large #1 } \end{center}}
\def\Author#1{\begin{center}{ \sc #1} \end{center}}
\def\Address#1{\begin{center}{ \it #1} \end{center}}

\newcommand\pubblock{\rightline{\begin{tabular}{l} \pubnumber\\
         \pubdate  \end{tabular}}}
\newenvironment{Abstract}{\begin{center}{\bf Abstract}\end{center} \bigskip \begin{quotation}  }{\end{quotation}}
\newenvironment{Presented}{\begin{quotation} \begin{center} 
             PRESENTED AT\end{center}\bigskip 
      \begin{center}\begin{large}}{\end{large}\end{center} \end{quotation}}

\def\beq{\begin{equation}}
\def\eeq#1{\label{#1}\end{equation}}
\def\eeqn{\end{equation}}

\def\beqa{\begin{eqnarray}}
\def\eeqa#1{\label{#1}\end{eqnarray}}
\def\eeqan{\end{eqnarray}}

\let\bar=\overbar

\def\Dslash{\not{\hbox{\kern-4pt $D$}}}
\def\dslash{\not{\hbox{\kern-2pt $\del$}}}

\def\msb{{\bar{\ssstyle M \kern -1pt S}}}

\begin{document}
\begin{titlepage}
\pubblock

\vfill


\Title{SUSY and high-$p_{T}$ flavor tagging at CMS}
\vfill
\Author{W. Kiesenhofer on behalf of the CMS Collaboration}  
\Address{Institute of High Energy Physics, Vienna, 1050, Austria}
\vfill


\begin{Abstract}
We present a result of a search for supersymmetry in final states with missing transverse energy and b-tagged jets. This search is performed with data collected by the CMS experiment at the LHC in pp-collisions at a center-of-mass energy of 7 TeV. Data-driven techniques used to measure the Standard Model background are demonstrated. The result is interpreted in terms of the constrained Minimal Supersymmetric Standard Model and compared to a similar search without any b-tag requirement.
\end{Abstract}

\vfill

\begin{Presented}
The Ninth International Conference on\\
Flavor Physics and CP Violation\\
(FPCP 2011)\\
Maale Hachamisha, Israel,  May 23--27, 2011
\end{Presented}
\vfill

\end{titlepage}
\def\thefootnote{\fnsymbol{footnote}}
\setcounter{footnote}{0}
%


\section{Introduction}

Searches for new physics beyond the Standard Model (SM) are commonly motivated by strong astrophysical evidence for dark matter, and by theoretical problems associated with explaining the observed particle masses while maintaining the mass hierarchies in the presence of quantum corrections~\cite{Witten:1981nf,Dimopoulos:1981zb}. 
Supersymmetry (SUSY)~\cite{Martin:1997ns,Wess:1974tw,Nilles:1983ge,Haber:1984rc,Barbieri:1982eh,Dawson:1983fw,Ellis:1983ew} is an extensively studied candidate for new physics with the potential to solve this problems. 
The SUSY particle spectrum contains new particles arising from a correspondence between SM fermions and SUSY partner bosons, as well as SM bosons and SUSY partner fermions.
At the Large Hadron Collider (LHC) supersymmetric particles, if they exist, are predicted to be predominantly produced via the fusion of two gluons into a:
\begin{itemize}
\item pair of gluinos (super-partners of the gluons),
\item pair of squarks (super-partners of the quarks),
\item gluino and a squark.
\end{itemize}
Gluinos and squarks decay via other supersymmetric particles into quarks and other SM particles until a lightest supersymmetric particle (LSP) is created. Under the assumption of R-parity conservation the LSP is stable and will generate missing energy transverse to the beam line (${\not} E_{T}$)  when it escapes the detector.
The details of SUSY decay chains are strongly model dependent and since we aim to design inclusive analysis, which are sensitive over a wide range of SUSY models, we do not consider them in our analysis.
However, for a large class of supersymmetric parameter sets squarks can be relatively light. If this is the case for sbottoms or stops, which can decay into b quarks, there may be an abundance of events with large ${\not} E_{T}$ and one or more b-quark jets.
This proceeding summarizes a search for events with two or more hadronic jets in the final state, 
significant transverse momentum imbalance, and at least one b-tagged jet~\cite{CMS:BJetCommissioning}.  It uses the full dataset collected by the CMS experiment~\cite{Adolphi:2008zzk} in 2010. This analysis extends a similar one without a b-tag requirement~\cite{Khachatryan:2011tk}. The momentum imbalance is characterized by the variable $\alpha_{T}$, definded in the next section.

The main backgrounds for this analysis arise due to standard model multi-jet production (QCD background), electroweak W and Z boson production (EWK) and top-quark pair production ($t\bar t$). Their estimation is discussed in section 3.

\section{The $\alpha_{T}$ variable}
The $\alpha_{T}$ variable charactarizes the momentum imbalance of jets in the transverse plane.
For a two jet system it is defined as
\begin{equation}\label{alphaT2jet}
\alpha_{T}=\frac{E^{j^{2}}_{T}}{M^{j^{1},j^{2}}_{T}}
\end{equation}
where $j^{2}$ is the jet with the lower transverse energy ($E_{T}$) and $M^{j^{1},j^{2}}_{T}$ is the invariant mass in the transverse plane of the two jets. Assuming massless jets, this can be rewritten for multi-jet events as 
\begin{equation}\label{alphaTmultijet}
\alpha_{T}=\frac12\frac{H_{T}-\Delta H_{T}}{\sqrt{H^{2}_{T}-{\not} H^{2}_{T}}}
\end{equation}
where ${\not} H_{T}=|\Sigma_{i} \vec p^{j^{i}}_{T}|$, $H_{T}=\Sigma_{i} p^{j^{i}}_{T}$ and $p_{T}$ denotes a jet's transverse momentum.
All jets considered in the analysis are grouped into two pseudo-jets such that $\Delta H_{T}=|p^{pseudojet1}_{T}-p^{pseudojet2}_{T}|$ becomes minimal.

The $\alpha_{T}$ variable is very effective in rejecting QCD background, which would otherwise dominate the signal selection. 
A well measured QCD event will result in $\alpha_{T}~0.5$. To account for finite jet energy and phi resolution a cut of $\alpha_{T}>0.55$ is used for the final event selection. A SUSY event with real missing transverse momentum due to the production of LSPs can exceed this value of $\alpha_{T}$ since ${\not}H_{T}$ can be sufficiently large compared to $H_{T}$.

\section{Event Selection}

Events that were considered in this analysis had to pass at least one trigger (based on $H_{T}$ at trigger level) during data taking. These triggers were measured to be over 99\% efficient for the final signal selection.
Jets were reconstructed using the anti-$k_{T}$~\cite{Cacciari:2008gp} algorithm and required to have $E_{T}>50$ GeV and $|\eta|<3$.
Since this analysis aims to be exclusively hadronic, events with an isolated lepton or  photon were vetoed.
In addition the final selection requires two jets with $E_{T}>100$ GeV, $|\eta|<2.5$ for the highest $E_{T}$ jet, $H_{T}>350$ GeV, at least one jet tagged as originating from a b quark, and $\alpha_{T}>0.55$.
To discriminate b jets from other flavors the TCHP (Track Counting High Purity) algorithm was used~\cite{CMS:BJetCommissioning}. For this discriminator three different working points are defined, namely a loose, a medium and a tight. Only the tight selection was used in the main analysis, whereas the others were used to collect control samples.

\section{Estimation of Backgrounds}

As stated previously the backgrounds for this analysis can be categorized into three groups: QCD, EWK and $t\bar t$.

The vast majority of events which belong to the QCD background do not feature large momentum imbalance and are therefore rejected by the $\alpha_{T }>0.55$ requirement. Sources of artificial momentum imbalance in QCD events, which can lift $\alpha_{T}$ over the signal requirement are jet under-measurement caused by inactive ECAL regions or multiple jets falling below the jet $E_{T}$ threshold of 50 GeV. Such events were identified and removed by control variables implemented for this purpose~\cite{CMS:RA1b}.

The EWK background processes (production of Z/W+jets) with real missing energy from the production of neutrinos are greatly suppressed by the requirement of at least one b jet. Thus $t\bar t$ becomes the dominant background in this analysis. 

\begin{figure}[htb]
\centering
\includegraphics[angle=0.0,width=0.49\textwidth]{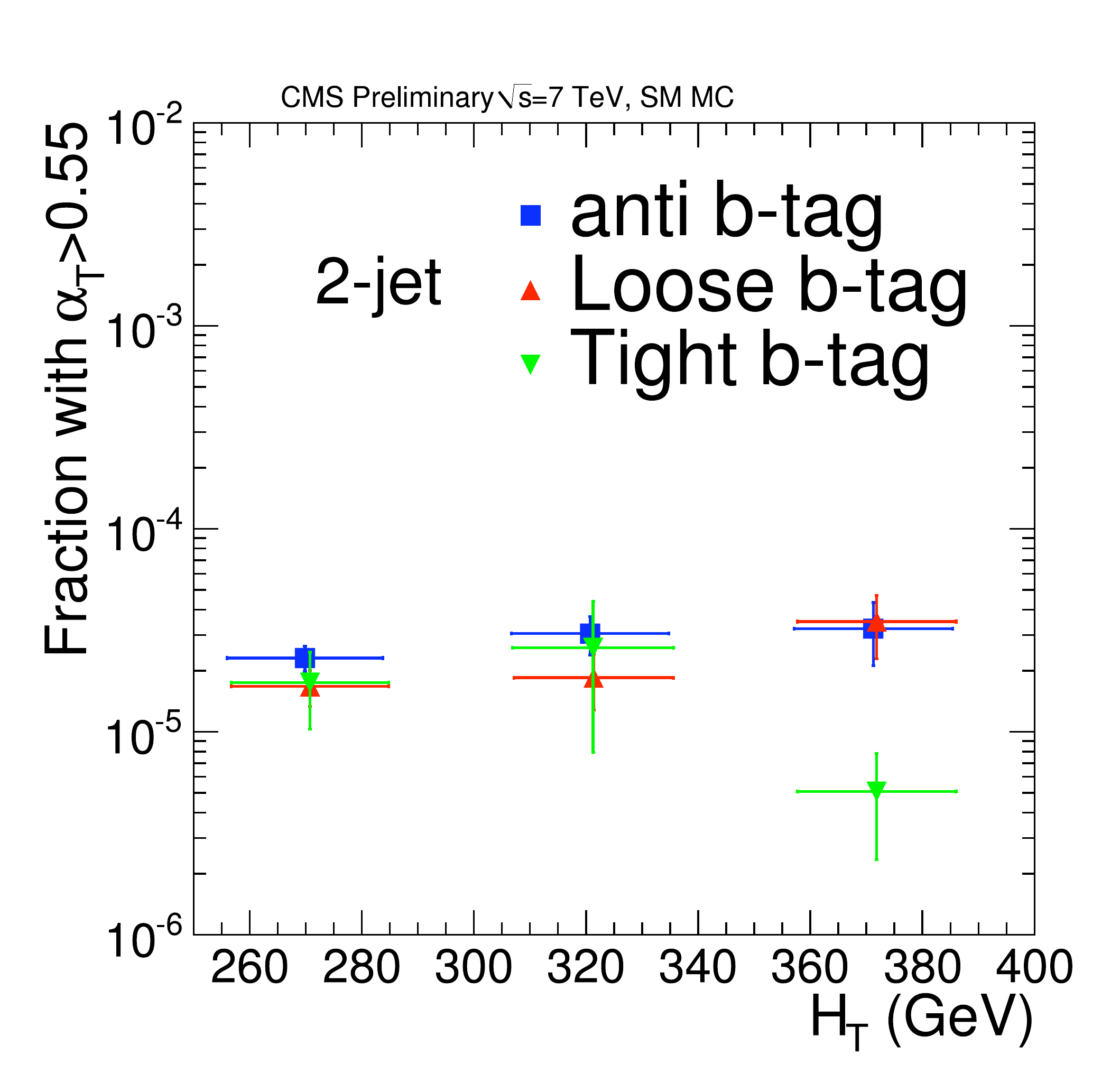}
\includegraphics[angle=0.0,width=0.49\textwidth]{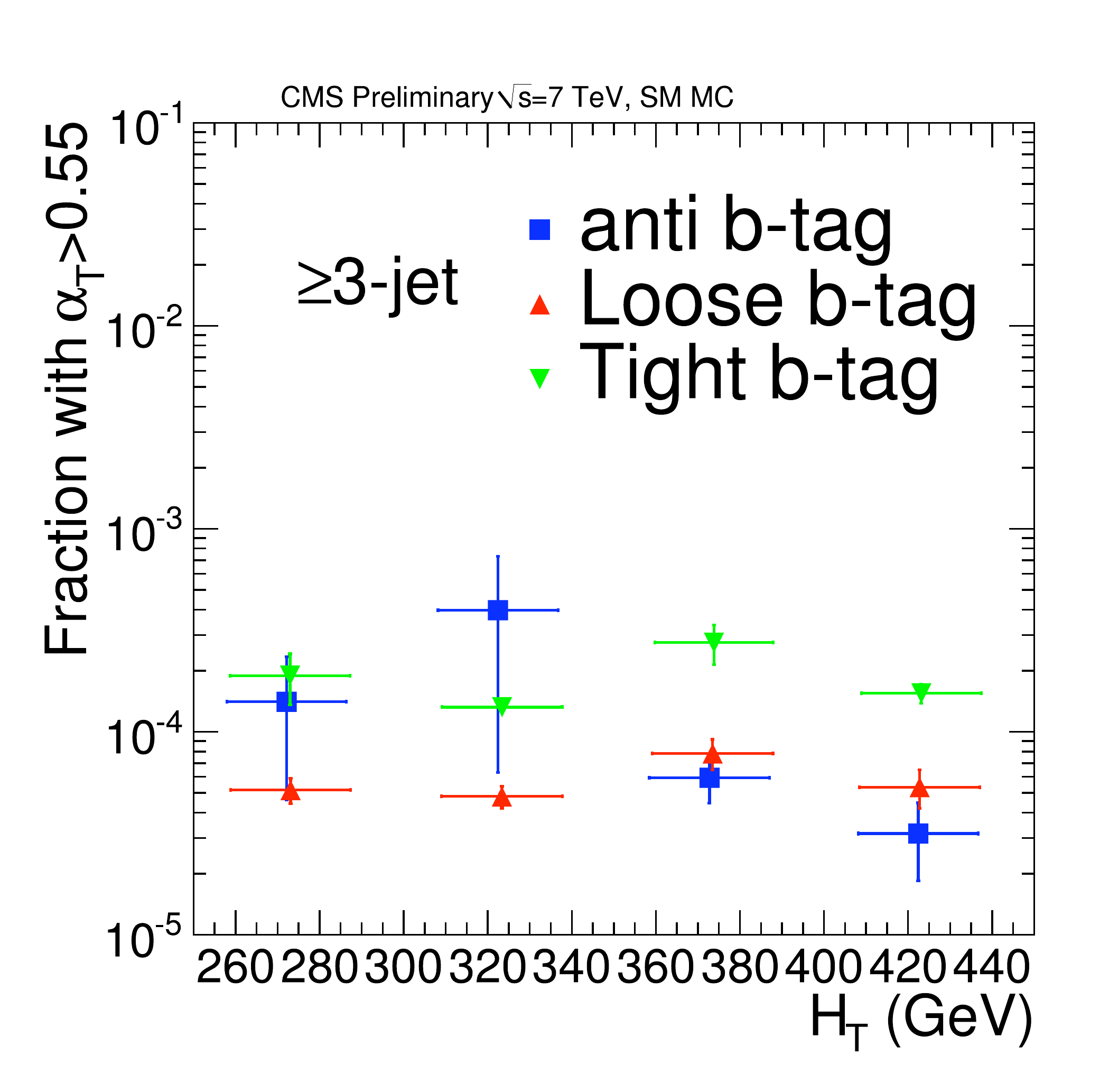}
\caption{$F(\alpha_{T} > 0.55)$ vs $H_{T}$ in SM simulation. In all cases, $F(\alpha_{T} > 0.55)$ is consistent with 
having no $H_{T}$ dependence.}
\label{fig:fig_alphaT_055}
\end{figure}

\begin{figure}[htb]
\centering
\includegraphics[angle=0.0,width=0.49\textwidth]{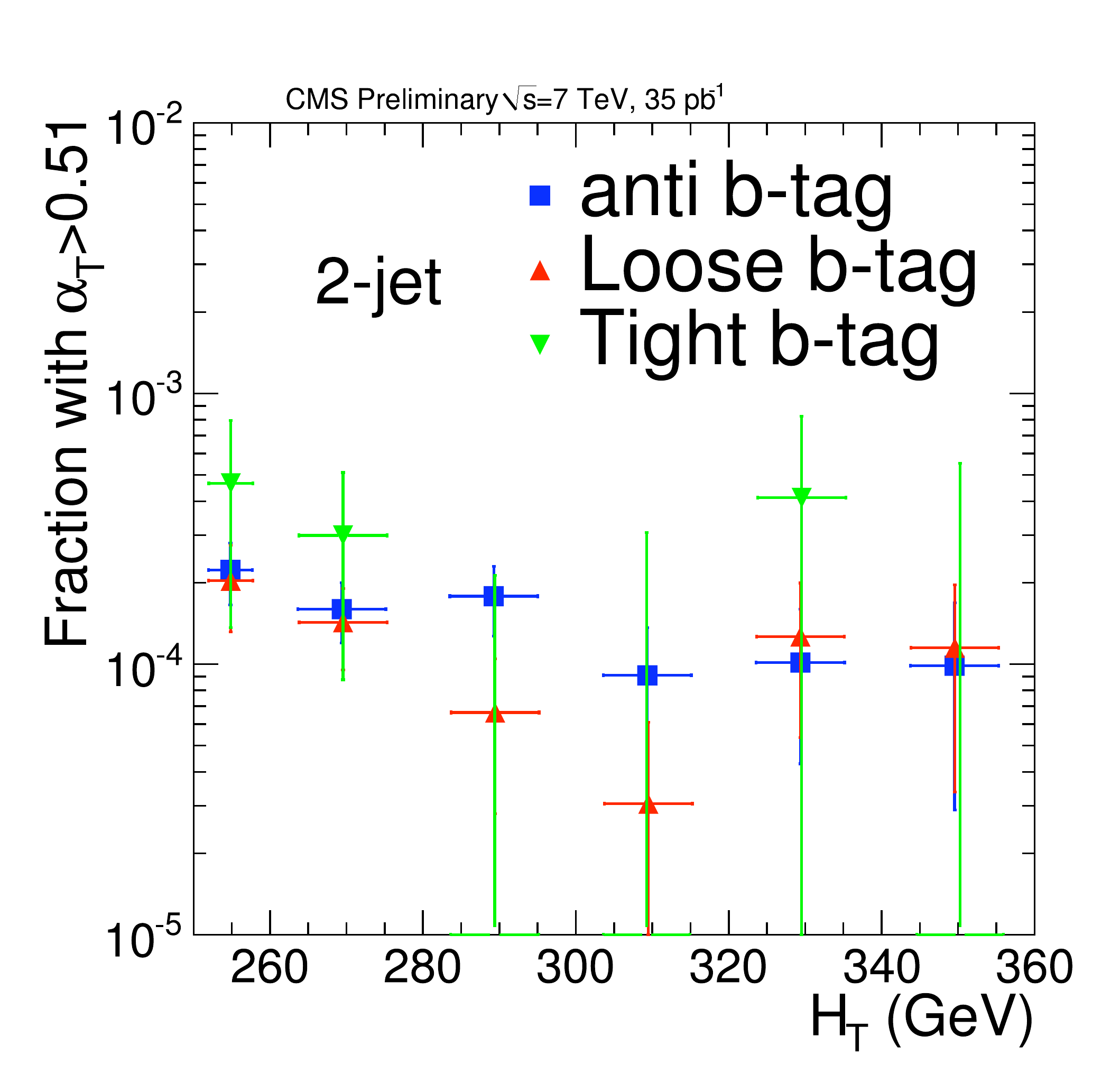}
\includegraphics[angle=0.0,width=0.49\textwidth]{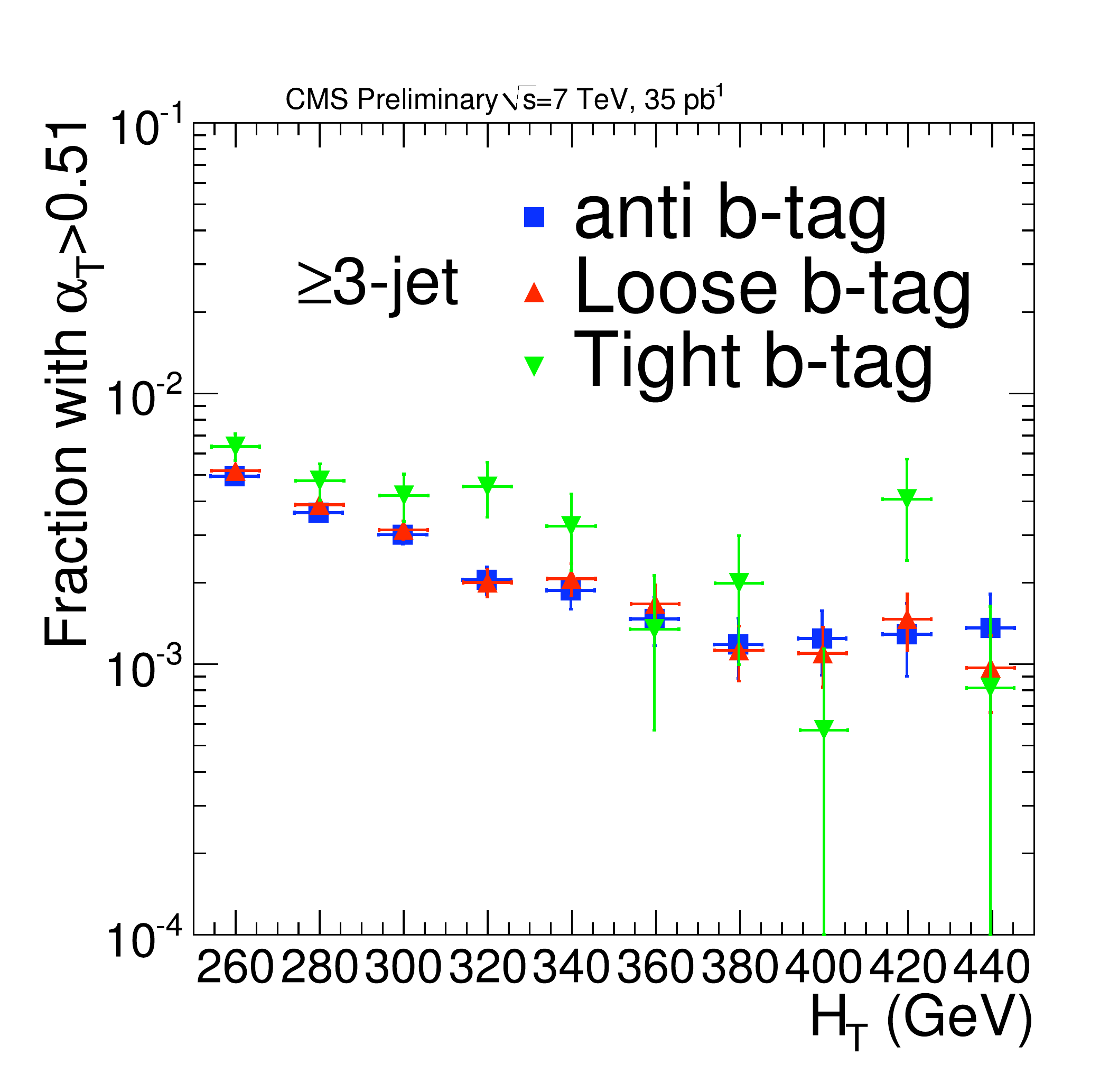}
\caption{$F(\alpha_{T} > 0.51)$ vs $H_{T}$ in data. $F(\alpha_{T} > 0.51)$ is a decreasing function of $H_{T}$ for lower $H_{T}$, 
indicating that QCD dominates these regions.}
\label{fig:fig_alphaT_051}
\end{figure}

Three independent estimation procedures have been employed to estimate the SM background in the final event selection. 
The main method is a data-driven procedure estimating all backgrounds simultaneously. In this method the fraction of all events with $\alpha_{T}>0.55$, denoted $F(\alpha_{T}>0.55)$, is measured in a lower $H_{T}$ control region and applied in the singal region (the jet $E_{T}$ threshold is scaled according to $H_{T}$). As studied extensively in Ref.~\cite{Khachatryan:2011tk} $F(\alpha_{T}>0.55)$ is independent of $H_{T}$ in samples were ${\not} H_{T}$ comes predominantly from real sources. This can be seen in Fig.~\ref{fig:fig_alphaT_055}, which shows $F(\alpha_{T}>0.55)$ vs $H_{T}$ in simulation of all SM processes combined.
In a QCD dominated sample $F(\alpha_{T}>0.55)$ is expected to be a decreasing function of $H_{T}$ due to the  $H_{T}$ dependence of the factors contributing to artificial ${\not} H_{T}$, such as jet energy resolution and jet $E_{T}$ threshold effects. Loosening the $\alpha_{T}$ requirement to $\alpha_{T}>0.51$ enriches the selection with QCD events and leads to decreasing $F(\alpha_{T}>0.51)$ as a function of $H_{T}$. This is shown in Fig.~\ref{fig:fig_alphaT_051}.
In data $F(\alpha_{T}>0.55)$ is consistent with having no $H_{T}$ dependence, indicating that EWK and $t\bar t$ backgrounds dominate. Also the anti-b-tagged data sample is consistent with having no $H_{T}$ dependence. Since the tight b-tag requirement only further suppresses the QCD background, the tight tagged sample is expected to have a negligible QCD contribution.
To predict the number of background events in the final signal region $F(\alpha_{T}>0.55)$ is measured in a lower $H_{T}$ control region (250-350 GeV) and multiplied by the number of events in the signal region before the $\alpha_{T}>0.55$ requirement. 
The results of this procedure are summarized in Table~\ref{tab:results}.

The two other background prediction methods estimate the $Z \rightarrow \nu \bar \nu$ and the $t\bar t$ contribution to the background and serve as a cross check to the first method.

To estimate $Z \rightarrow \nu \bar \nu$, which is expected to be the dominant EWK background, a sample of  $Z \rightarrow \mu^{+}\mu^{-}$ events with  $\geq2$ jets is selected and the fraction of events with a b-tagged jet is measured. In a second step a sample is selected with no b-tag requirement and slightly looser selection requirements than the final signal selection. The number of events in this sample is scaled by the measured b-tag fraction, corrected for the muon identification efficiency and acceptance and multiplied by $\frac{BR(Z \rightarrow \nu \bar \nu)}{BR(Z \rightarrow \mu^{+}\mu^{-})}\approx6$. The result of this procedure shows a slight over-prediction of the number of  $Z \rightarrow \nu \bar \nu$ events due to looser selection cuts, but is consistent with the prediction of the former method and with the prediction from simulation.
 
Since $t\bar t$ is expected to be the dominant background it is important to cross check its contribution with a second estimation method.
Simulation studies indicate that 72\% of the $t\bar t$ background consists of events with hadronic tau decays.
To estimate the hadronic tau decay yield, events with one or two muons are selected and used to emulate the hadronic decays of taus. For each muon, the presence of a jet from the hadronic decay of a tau lepton is emulated. The jet $E_{T}$ is determined as fraction of the muon $p_{T}$, where the fraction is drawn from a distribution extracted from simulation. Again a less stringent version  of the final event selection is used in which the $E_{T}$ threshold of the leading two 
jets is 80 GeV, the $H_{T}$ requirement is 280 GeV, and a medium b-tag is required. The measured value of $F(\alpha_{T} > 0.55)$ in this sample is multiplied 
by the total number of emulated events in the signal region with no $\alpha_{T}$ requirement. This value is corrected for the muon selection efficiency and acceptance and the hadronic tau decay branching ratio to obtain the hadronic tau decay yield. To account for the entire $t\bar t$ background, 
the predicted hadronic tau decay yield is scaled by a simulation derived factor. A 50\% systematic uncertainty is assigned for this factor.
This procedure again yields a slight over prediction in $t\bar t$ simulation.

This two cross checks confirm that the total background is small and consistent with the inclusive prediction. 

\begin{table}[!hbtp]
\begin{center}
\begin{tabular}{l|cccc}  
\hline\hline
N-jets      &  MC                           &  Background Prediction                                   & Data & LM0                    \\ \hline
 $\geq2$  &   $1.61 \pm 0.26$  &   $0.33^{+0.43}_{-0.33}$(stat)$\pm0.13$ & 1       & $14.2\pm0.3$   \\
\hline\hline
\end{tabular}
\caption{Predicted and observed events for 35 $pb^{-1}$. The prediction comes from the $\alpha_{T}$ vs $H_{T}$ extrapolation described in Section 3. The LM0 uncertainty is statistical only.}
\label{tab:results}
\end{center}
\end{table}

\section{Results}

\begin{figure}[htb]
\centering
\includegraphics[width=0.9\textwidth]{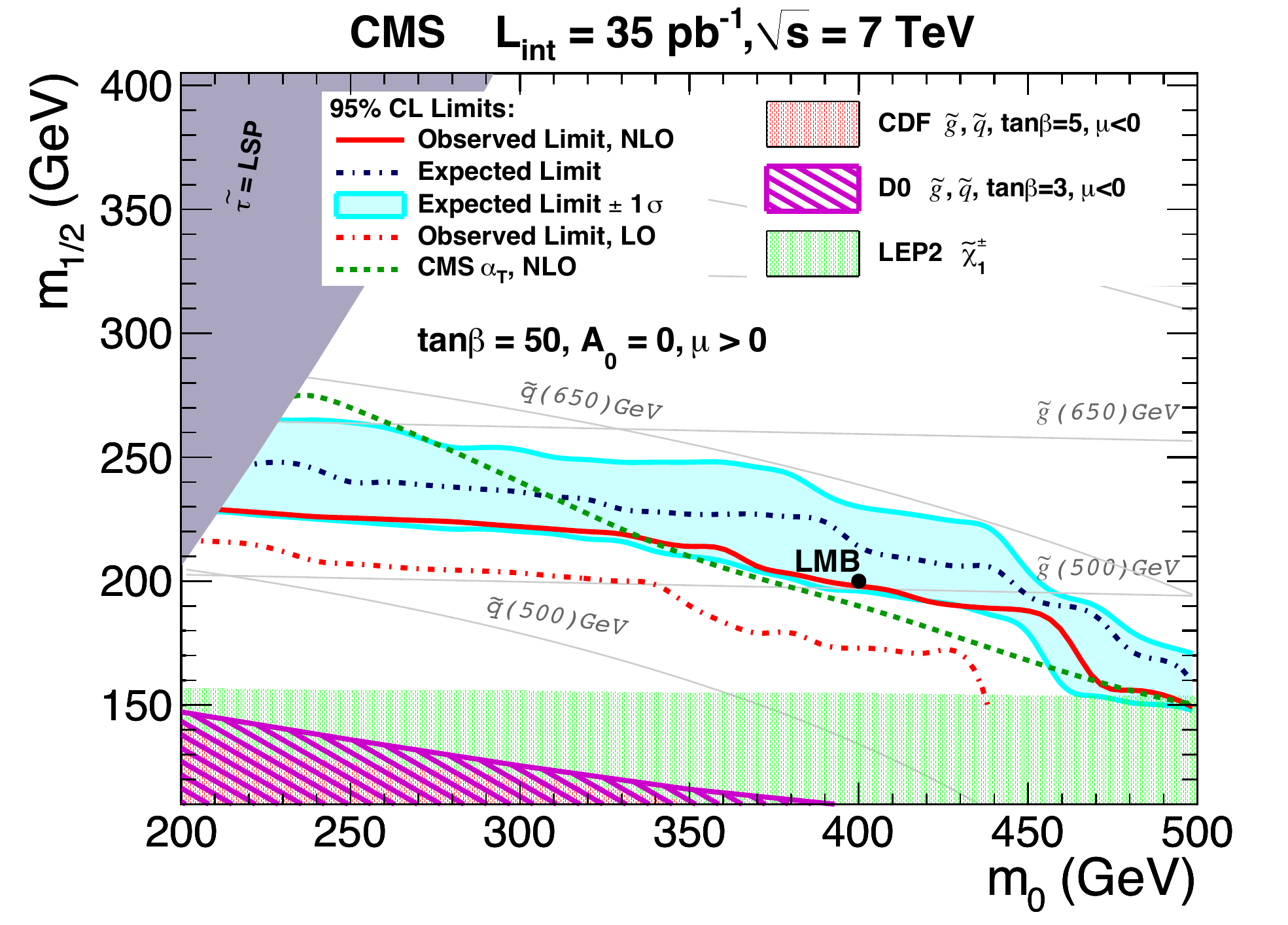}
\caption{Exclusion regions in the ($m_{1/2}$, $m_{0}$) plane for one set of CMSSM parameters, for this 
analysis (red), and the non-b tagged version~\cite{Khachatryan:2011tk} (green).}
\label{fig:fig_exclusion}
\end{figure}

The observation of 1 data event in the signal region is consistent with the background predictions summarized in the last section. Together with an estimated systematic uncertainty of 24\% on the signal selection efficiency~\cite{Chatrchyan:2011bj} this allows the determination of a 95\% confidence level upper limit on the predicted number of observed signal events $N^{obs}_{95}=4.7$.
Figure~\ref{fig:fig_exclusion} shows the excluded region in the ($m_{1/2}$, $m_{0}$) plane for CMSSM (constrained Minimal Supersymmetric Standard Model)~\cite{Kane:1993td} parameters $A_{0} = 0$ GeV, 
$\tan\beta = 50$, and $\mu > 0$. The excluded region is extended with respect to that of Ref.~\cite{Khachatryan:2011tk} without 
b tagging, also shown, values of $m_{0}$ above 350 GeV, where b production is frequent. For models 
with infrequent b production, Ref.~\cite{Khachatryan:2011tk} sets more stringent limits, whereas this analysis has 
greater sensitivity to models with frequent b production.

\section{Conclusion}

In this proceeding we present a search for events with multiple jets, at least one of which must be b tagged, and signiﬁcant 
transverse momentum imbalance, using the \mbox{35 $pb^{-1}$} of the CMS 2010 dataset. No evidence for new physics is observed. The result of the search is characterized as an exclusion region in SUSY parameter space and compared to an equivalent analysis without b tagging.

\end{document}